\documentclass[
 reprint,
superscriptaddress,
showkeys,
amsmath,amssymb,
aps,prb,
floatfix,
]{revtex4-1}

\usepackage{graphicx}
\usepackage{dcolumn}
\usepackage{bm}
\usepackage{physics}
\usepackage{natbib}

\begin{document}

\preprint{APS/123-QED}

\title{Real-space imaging and flux noise spectroscopy of magnetic dynamics in Ho$_2$Ti$_2$O$_7$}

\author{Christopher A. Watson}
\thanks{These two authors contributed equally}
\affiliation{
 Stanford Institute for Materials and Energy Sciences,
SLAC National Accelerator Laboratory,\\
2575 Sand Hill Road, Menlo Park, CA 94025, USA
}
\author{Ilya Sochnikov}
\thanks{These two authors contributed equally}
\affiliation{
Geballe Laboratory for Advanced Materials, Stanford University, Stanford, California 94305, USA}
\affiliation{
Department of Physics, University of Connecticut, Storrs, Connecticut 06269, USA}
\author{John R. Kirtley}
\affiliation{
Geballe Laboratory for Advanced Materials, Stanford University, Stanford, California 94305, USA}
\author{Robert J. Cava}
\affiliation{
Department of Chemistry, Princeton University, Princeton, New Jersey 08540, USA}
\author{Kathryn A. Moler}
\affiliation{
 Stanford Institute for Materials and Energy Sciences,
SLAC National Accelerator Laboratory,\\
2575 Sand Hill Road, Menlo Park, CA 94025, USA
}
\affiliation{
Geballe Laboratory for Advanced Materials, Stanford University, Stanford, California 94305, USA}

\date{\today}

\begin{abstract}
Holmium titanate (Ho$_2$Ti$_2$O$_7$) is a rare earth pyrochlore and a canonical example of a classical spin ice material.  Despite the success of magnetic monopole models, a full understanding of the energetics and relaxation rates in this material has remained elusive, while recent studies have shown that defects play a central role in the magnetic dynamics.  We used a scanning superconducting quantum interference device (SQUID) microscope to study the spatial and temporal magnetic fluctuations in three regions with different defect densities from a Ho$_2$Ti$_2$O$_7$ single crystal as a function of temperature.  We found that the magnetic flux noise power spectra are not determined by simple thermally-activated behavior and observed evidence of magnetic screening that is qualitatively consistent with Debye-like screening due to a dilute gas of low-mobility magnetic monopoles.  This work establishes magnetic flux spectroscopy as a powerful tool for studying materials with complex magnetic dynamics, including frustrated correlated spin systems.
\end{abstract}


\maketitle

\section{\label{sec:HTOIntro}Introduction}
Classical spin ices such as holmium titanate (Ho$_2$Ti$_2$O$_7$) have generated intense interest, both theoretical and experimental, in the last two decades.\cite{denHertog00, Bramwell01, Fennell09, Morris09, Bramwell09, Jaubert09, Giblin11, Castelnovo, Castelnovo12, Castelnovo15, Kassner15} 
The pyrochlore lattice of corner-sharing tetrahedra hosts a magnetic holmium atom at each shared vertex, and the local crystal field environment causes each holmium moment to point directly into one or the other of the adjacent tetrahedra.  The term ``spin ice'' comes from the analogy between the ground state manifold, which has two spins pointing in and two spins pointing out of each tetrahedron (the \emph{ice rule}), and the freezing of water ice, in which the two electron lone pairs on the oxygen atom line up with the hydrogen atoms of adjacent water molecules.\cite{denHertog00}

A single spin flip from the ground state manifold results in an excitation that is dipole-like.  A subsequent flip of a nearest-neighbor spin can restore the ice rule in one of the tetrahedra while violating it in the adjacent tetrahedron, extending the dipole to next nearest tetrahedra.  Continuing this process, one end of the dipole can be taken away entirely, leaving a single monopole-like excitation behind.\cite{Castelnovo}  Much of the recent theoretical work has focused on these emergent magnetic monopole models.\cite{Balents10,Castelnovo08,Castelnovo,Castelnovo12,Jaubert09,Jaubert11,Fennell09,Bramwell09} Nevertheless, it has subsequently become clear that defects, such as oxygen vacancies and stuffed spins (additional spins from Ho atoms occupying Ti sites), must be accounted for in understanding the full magnetic dynamics of spin ice.\cite{Revell13,Sala14,Baroudi15}  Progress in understanding these dynamics has been limited by a relative lack of suitable tools for microscopic magnetic studies.

In this paper, we demonstrate the utility of magnetic flux noise spectroscopy as such a tool for studying frustrated correlated spin systems.  We used a scanning superconducting quantum interference device (SQUID) microscope with a gradiometric SQUID magnetometer that has been described previously.\cite{Huber08}  With a spatial resolution of $4.6~\mu$m and a magnetic flux noise floor of order $1~\mu\Phi_0/\sqrt\textrm{Hz}$, we measured the temperature dependence of spatial magnetic correlations and of the magnetic flux noise spectrum in three regions with different defect densities from a Ho$_2$Ti$_2$O$_7$ single crystal.  We observed qualitative deviations from simple thermally activated behavior, which would predict a Lorentzian noise spectrum with a characteristic time that follows an Arrhenius law in temperature.  Furthermore, we found evidence of screening at low frequencies and high temperatures, which we compare to a model for Debye-like screening from the theoretically predicted gas of magnetic monopoles.\cite{Bramwell,Ryzhkin} 

\section{\label{sec:HTOSetup}Experimental Setup}
\begin{figure*}
\centering
\includegraphics[width=\linewidth]{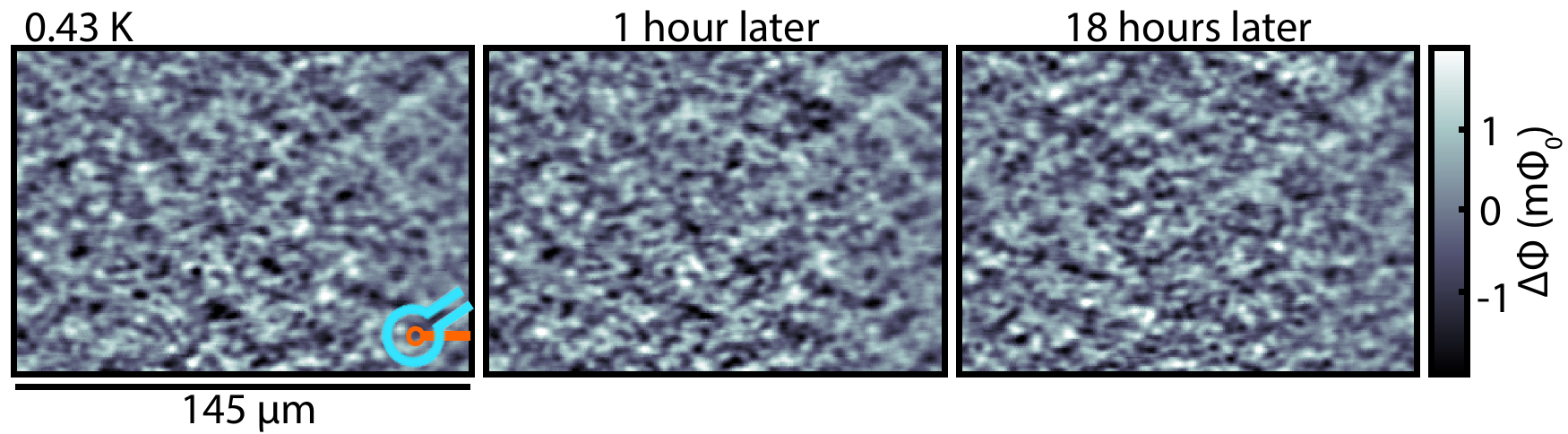}
\caption[Magnetometry scans at 430 mK taken over 18 hours]{\label{timeseries}Magnetometry scans of Sample A at 430 mK taken over 18 hours. Overlay in first panel is scale drawing of SQUID pickup loop (orange) and field coil (blue).  The pickup loop size sets the spatial resolution of the images; resolution-limited features in the scans are qualitatively similar from one scan to the next, but are different in the details.  This demonstrates that magnetic dynamics persist, even over long timescales and at the lowest measured temperatures, and that the magnetic texture fails to order or otherwise converge.}
\end{figure*}
We measured two samples from a single crystal grown via the floating-zone method which has previously been characterized elsewhere;\cite{Baroudi15} sample A is from near the center of the growth boule while Sample B is from near the edge, where the density of defects was somewhat higher.  Sample A was transparent pink, while Sample B was cloudy but translucent pink, with a dark, opaque region at one corner.  We measured both regions of Sample B, which we will subsequently refer to as Samples B1 and B2, respectively. We fractured sample A to obtain a smooth but not flat surface with roughly [111] orientation.  For Sample B, we prepared a polished [111] surface with $< 1~\mu$m grit polishing film and isopropanol.

For two-dimensional image data, we defined the scan surface by determining the height at which the SQUID was in contact with the sample at a series of locations and fitting a two-dimensional, second order polynomial to the surface topography.  We rastered the SQUID parallel to this surface at a nominal height of 1 $\mu$m. For one dimensional scans, we acquired a single row of the image repeatedly to discern the time evolution.

We obtained magnetic flux noise power spectra by placing the SQUID in contact with the sample and collecting each spectrum with an SR760 FFT Spectrum Analyzer in three overlapping segments (3.8 mHz--1.52 Hz, 488 mHz--195 Hz, and 125 Hz--49.9 kHz), using a Hanning window function and 128 exponentially weighted averages.  

\section{\label{sec:HTOResults}Results}
We present magnetometry scans of Sample A taken at our base temperature of 430 mK in Fig.~\ref{timeseries}.  At this temperature, features which are limited by the spatial resolution of the SQUID magnetometer (4.6 $\mu$m) dominate each scan, suggesting that the dynamics are slow compared to the scan speed. Repeated scans appear as different panels, at times indicated at the top of each panel, showing long timescale fluctuations despite qualitative similarity from scan to scan.  As expected for a truly frustrated spin system, the sample fails to show convergence or ordering of the magnetic texture even over the hold time of our cryostat, in excess of two days, at 430 mK. 

\begin{figure}
\centering
\includegraphics[width=\columnwidth]{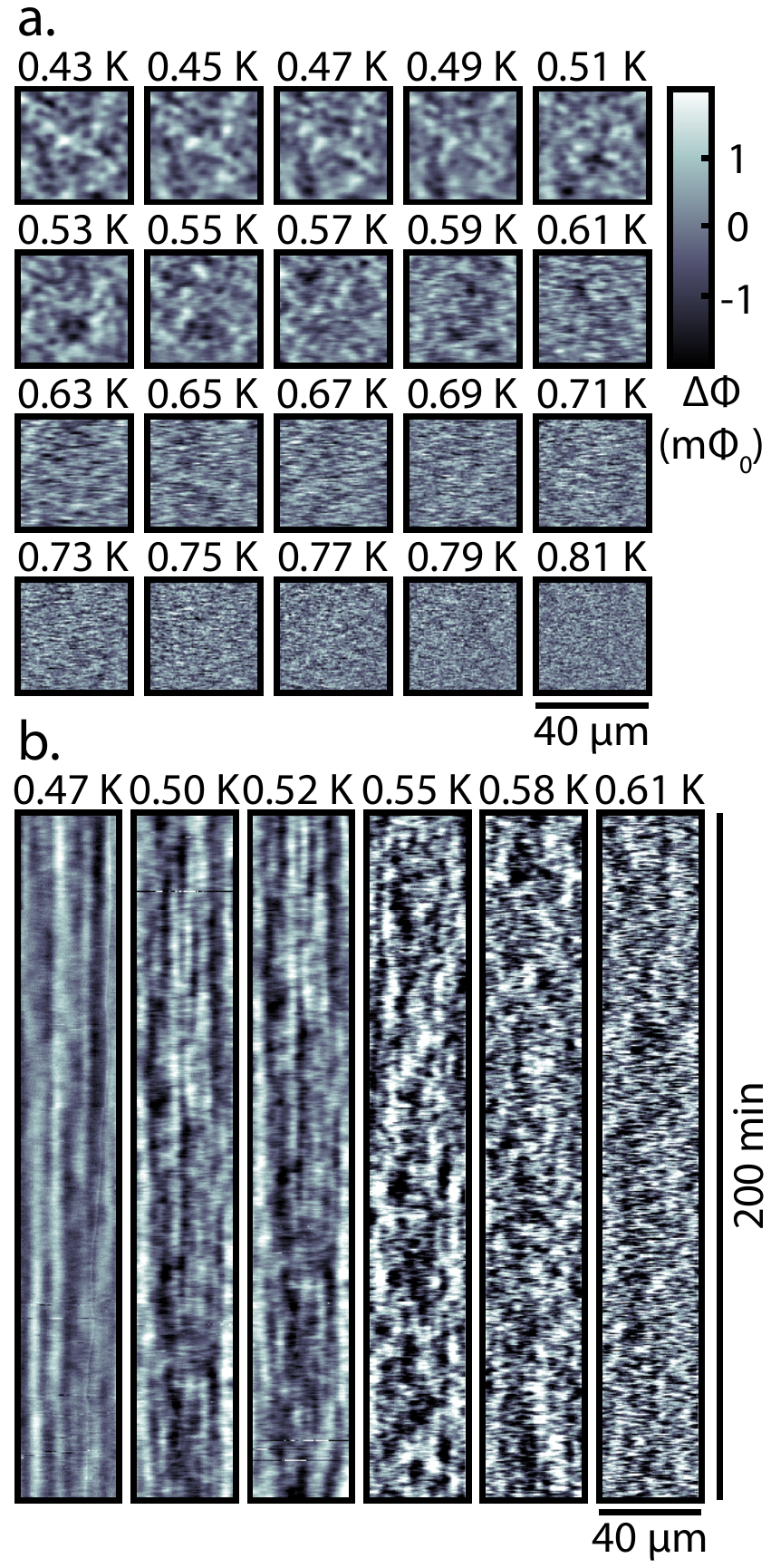}
\caption[Magnetometry scans as a function of temperature]{\label{HTOtempseries}Magnetometry scans of Sample A as a function of temperature. (a) Two-dimensional scans, as in Fig.~\ref{timeseries}, from base temperature to 810 mK. As the temperature increases, the observed fluctuations become sub-resolution, suggesting that there are temporal fluctuations that are fast compared to the scan speed. (b) One-dimensional scans vs. time from 470 mK to 610 mK.  Each series is 200 minutes long, and the vertical correlations of pixels from row to row characterize the correlation time, which is of order hours at 470 mK but falls to seconds by 610 mK.}
\end{figure}

In magnetometry scans conducted as a function of temperature, from base temperature to 810 mK, we observe that magnetic dynamics quicken as the temperature is increased.  In Fig.~\ref{HTOtempseries}, we show both two-dimensional scans [Fig.~\ref{HTOtempseries}(a)], each acquired over several minutes, and one-dimensional scans as a function of time [Fig.~\ref{HTOtempseries}(b)], each acquired over 200 minutes with each row taking approximately 12 seconds, taken at various temperatures.  As in Fig.~\ref{timeseries}, scans at the lowest temperatures (first panels in each part of the figure) show resolution-limited features, implying that temporal dynamics occur on timescales long compared to the sampling rate. The first panel in Fig.~\ref{HTOtempseries}(a) shows this explicitly, as the features persist over many rows, suggesting a correlation time of order hours.  As the temperature is increased, the features in the scans in Fig.~\ref{HTOtempseries}(a) vary on shorter length scales, indicating that there are faster temporal variations coming into play.  This is made manifest by comparing adjacent rows in the various panels of Fig.~\ref{HTOtempseries}(b), where the correlation time falls to order seconds by 610 mK.  Due to the limited scan speed of the SQUID, these measurements cannot resolve magnetic dynamics at temperatures above 1 K, as it is difficult to unambiguously distinguish spatial and temporal variations. To measure at higher temperatures, we instead fix the position of the SQUID in contact with the sample surface and measure the magnetic flux as a function of time only.  By taking the Fourier transform of time series data, we obtain a magnetic flux noise power spectrum.

The key results of this paper are contained in plots of the natural logarithm of the magnetic flux noise power spectra, in units of $\Phi_0^2/\textrm{Hz}$, as a function of the standard logarithm of frequency, $\log(f)$, and the inverse temperature, $1/T$.  Were the sample an ensemble of identical but non-interacting, thermally-excited Ising spins, we would observe a Lorentzian noise spectrum, $S_\Phi = c\tau/(1+\omega^2\tau^2)$, with $c$ constant, $\omega$ the measurement frequency, and $\tau$ a temperature-dependent characteristic time. The thermal excitation over the Ising barrier would yield an Arrhenius law for that temperature dependence, $\tau = \tau_0e^{E_a/k_BT}$, where $\tau_0$ is a microscopic attempt time, $E_a$ is an activation energy, $k_B$ is the Boltzmann constant, and $T$ is the temperature.  In this illustrative example, the contours of constant noise power would be vertical at high temperatures and linear at low temperatures. For Lorentzian noise spectra, the noise power monotonically increases for decreasing frequency, down to a characteristic frequency at which it plateaus.  The line formed by the maxima of horizontal line cuts gives the Arrhenius law: the slope gives $E_a$ while the intercept is related to $\tau_0$.  

\begin{figure}
\centering
\includegraphics[width=\columnwidth]{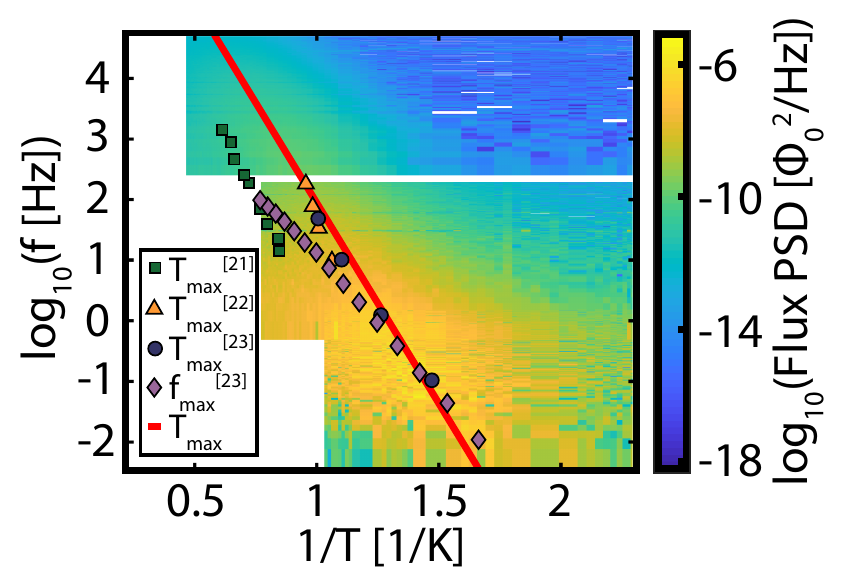}
\caption{\label{pastdata}Magnetic flux noise power spectra as a function of temperature for Sample A.  Overlaid red line is a linear fit to maxima for each row, while markers indicate ac susceptibility data from previous works for comparison.}
\end{figure}

In Fig.~\ref{pastdata}, we show the magnetic flux noise spectra for Sample A from 430 mK to 4 K. The data are manifestly non-Arrhenius: the contours of constant noise power at low temperatures flare out, with the noise power from 1--100 Hz at 430 mK higher than would be expected for a Lorentzian following an Arrhenius law.  For temperatures above 650 mK, the flux noise power as a function of decreasing frequency reaches a maximum and then falls off sharply at lower frequencies, below 1 Hz at 1 K.  This feature suggests that, regardless of whether the observed magnetic flux noise spectra result from the dynamics of magnetic monopoles or some other microscopic origin, there is a source of magnetic screening within the sample. The overlaid red line in Fig.~\ref{pastdata} is a best fit line (Arrhenius law) to the maxima of each row.  Comparing the extracted Arrhenius law to previously reported bulk ac susceptibility data,\cite{Ehlers03,Matsuhira00,Quilliam11}  we find that it is in close agreement, suggesting that we are measuring the same magnetic dynamics as have previously been reported.\cite{Matsuhira00, Snyder01, Matsuhira01, Ehlers03, Snyder04, Quilliam11, Matsuhira11, Jaubert11, Yaraskavitch12, Takatsu13, Bovo13, Paulsen14}

\begin{figure}
\includegraphics[width=\columnwidth]{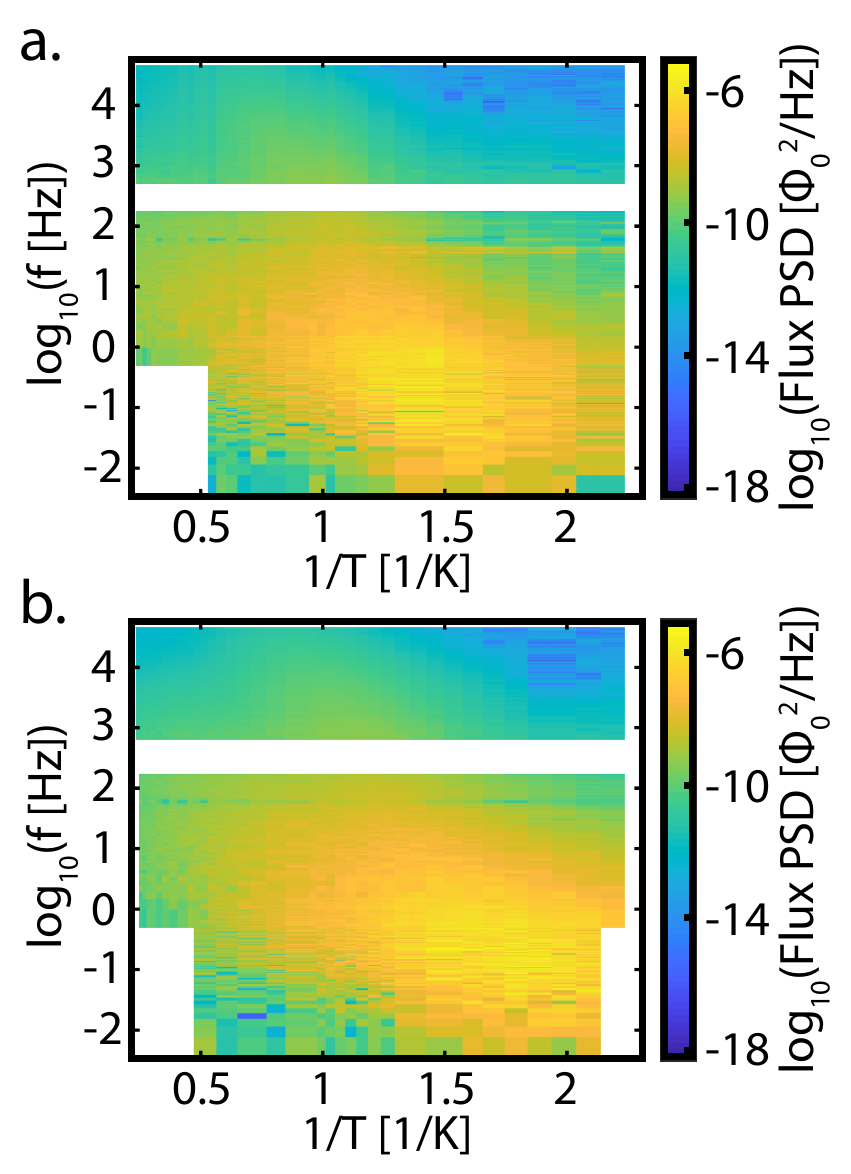}
\caption{\label{defectseries}Comparison of magnetic flux noise power spectra from two locations on second sample (a.  B1; b. B2), demonstrating different defect densities. The central, Arrhenius-like feature seen in Sample A is also seen in both data sets; the feature becomes broader, more diffuse, and more shallow as the defect density is increased, consistent with the barrier to spin flips shortening and becoming broadened with increased magnetic disorder.  Qualitative deviations from Arrhenius behavior, seen as excess noise at center left and bottom right of each panel, and screening at bottom left of each panel, appear in all data sets and are quantitatively similar in their frequency and temperature dependence.  These comparisons suggest that the central, Arrhenius-like feature is affected by magnetic defects, while the deviations from Arrhenius behavior are universal.}
\end{figure}

To understand the impact on the magnetic dynamics of defects, such as those introduced by additional magnetic holmium atoms on titanium sites (stuffed spins),\cite{Lau06, Sala14, Sen14, Baroudi15} we measured two regions from an additional sample (Sample B) taken from nearer to the edge of the growth boule.  The additional flux noise spectra are shown in Fig.~\ref{defectseries}, together with those from Sample A.  The predominant Arrhenius-like feature smears out considerably and becomes somewhat less steep as the defect density is increased, implying a broadening distribution of activation energies that are lower on average.  This is consistent with expectations for increased defect densities, as magnetic disorder broadens and reduces the barrier for individual spins in the sample to flip.

The qualitative deviations from Arrhenius behavior seen in Sample A can be seen more clearly in Sample B.  The flaring out of the contours of constant noise power at low temperatures have corresponding features at high temperatures, seen above 1.3 K from 0.1--100 Hz.  The correspondence of the excess noise features at the highest and lowest temperatures suggests that there is an additional source of magnetic dynamics in these samples at lower frequencies than has been accessible in previous ac susceptibility measurements.  Furthermore, we see that the screening behavior in Sample A is also seen in Sample B, and that it is quantitatively comparable across all three samples, independent of the defect density. 

\section{\label{sec:HTODiscussion}Discussion}
Having shown the flux noise spectra from our samples, we now turn to the question of what we expect from a dilute gas of monopoles. The basic form of our model is a Lorentzian noise term (this form for the noise due to magnetic monopoles was previously robustly justified by Ryzhkin\cite{Ryzhkin}), modified by a Debye-like screening term:
\begin{equation}
S_{\Phi} = C\left(\frac{\tau_{Mon}}{1+\omega^2\tau_{Mon}^2}\right)\frac{\omega^2/\omega_c^2}{1+\omega^2/\omega_c^2}
\end{equation}
where $C$ is an overall scaling constant, $\omega$ is the angular frequency, $\omega_c$ is a characteristic cutoff frequency for Debye screening as described below, and  $\tau_{Mon}$ is the characteristic time associated with spin relaxation, $\tau_{Mon} = \tau_{Mon,0}/x(T)$, where $\tau_{Mon,0}$ is the microscopic monopole hopping time and $x(T)$ is the temperature-dependent monopole concentration.  This relaxation time is responsible for monopole hopping by way of the fluctuation-dissipation theorem.\cite{Bramwell}

The Debye-H\"uckel model, as applied to the case of magnetic monopoles in spin ice by Castelnovo, Moessner, and Sondhi,\cite{Castelnovo} implies that the magnetic fields due to a source magnetic charge will be screened by a cloud of magnetic monopoles equal and opposite in charge to the source.  This screening occurs over a length scale, the Debye length $l_D$, which is given by:
\begin{equation}
l_D = \sqrt{\frac{k_BTV_0}{\mu_0Q^2x}}
\end{equation}
with $T$ the temperature, $V_0$ the volume of the diamond lattice site, $Q$ the monopole charge, and $x$ the monopole concentration.  The Debye-H\"uckel concentration for monopoles, $x(T)$, can be calculated iteratively as described in Ref.~\cite{Castelnovo}.

The Debye length is of order $50$ nm for the lowest temperatures at which we performed flux spectroscopy, and monotonically decreases as the temperature rises, such that it is always far smaller than the spatial resolution for the SQUID.  This suggests that in the presence of monopoles there would be no observable magnetic fluctuations whatsoever; however, because the monopoles have a finite mobility, only slowly varying magnetic fields are screened.

The mobility, $\mu$, appears in the characteristic Debye frequency by way of the diffusivity, $D$, and the Einstein relation:
\begin{equation}
\omega_c = \frac{D}{l_D^2} = \frac{\mu k_BT}{l_D^2}.
\end{equation}
The mobility itself has been calculated from Monte Carlo simulations:\cite{Castelnovo}
\begin{equation}
\mu = \frac{4}{27}\frac{a_d^2}{k_BT\tau}
\end{equation}
where $a_d$ is the lattice constant and $\tau$ is the Monte Carlo step time, which we identify as the same magnetic relaxation time as is used in the noise calculation above. 

Since the Debye-H\"uckel concentration can be computed directly, the noise and screening due to magnetic monopoles can be modeled with only two free parameters, the microscopic monopole hopping time and an overall scale factor that takes into account geometric factors that couple the fluctuating population of monopoles and the SQUID magnetometer.

\begin{figure}
\centering
\includegraphics[width=\columnwidth]{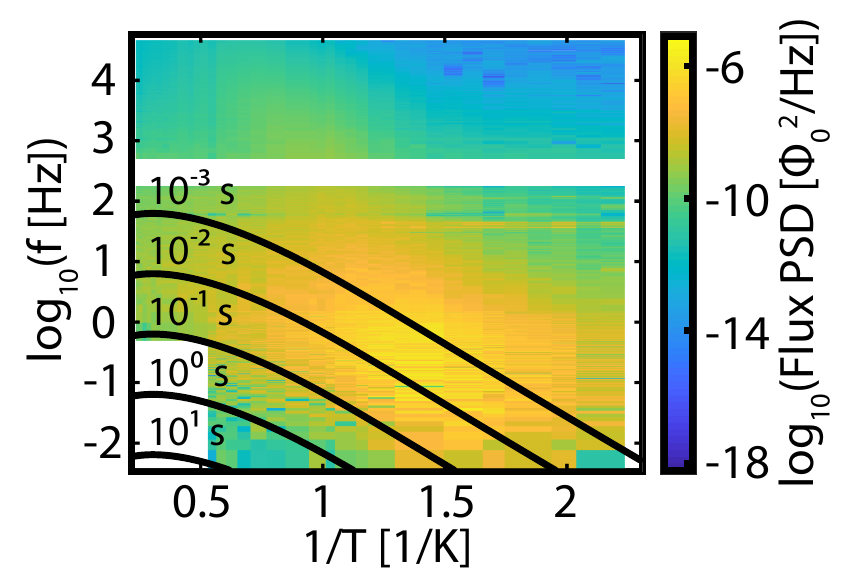}
\caption{\label{screeningcountours}Temperature dependence of the Debye screening critical frequency, $\omega_c$, for different values of the monopole hopping time. The contours indicate $\omega_c(T)$ for each hopping time, as labeled.}
\end{figure}

In Fig.~\ref{screeningcountours}, we overlay contours that give the characteristic frequency as a function of temperature, $\omega_c(T)$, for the Debye screening for various values of the monopole hopping time, as indicated.  Noting that the characteristic frequency is where the screening is of order unity, while the blue region in the bottom left corner of Fig.~\ref{screeningcountours} is where the noise is already reduced by orders of magnitude, we identify the monopole hopping time as 1--10 ms, in agreement with some previous measurements.\cite{Snyder04, Jaubert09, Quilliam11, Kaiser15, Tomasello15}  We also note the qualitative agreement between the data and plotted contours for the temperature dependence of the screening.

\begin{figure}
\centering
\includegraphics[width=\columnwidth]{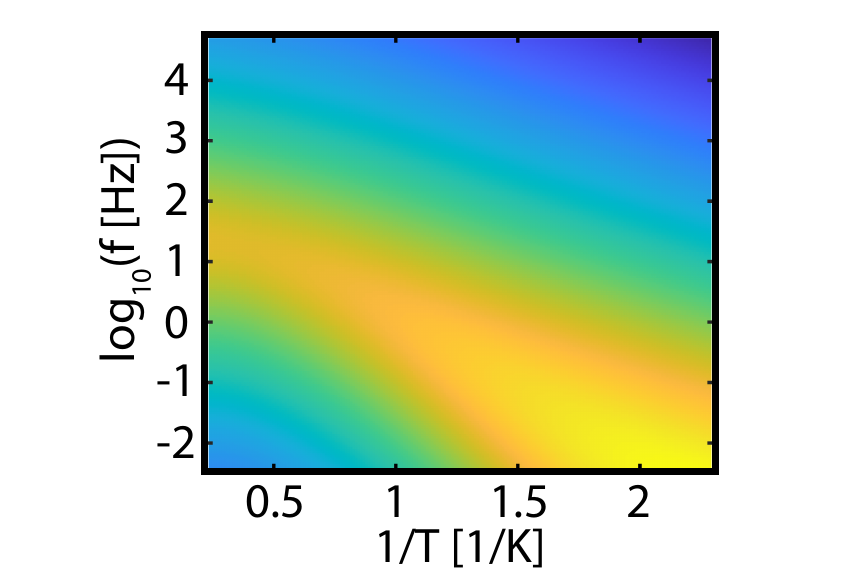}
\caption{\label{monomodel}A model of the expected magnetic dynamics for monopoles in Ho$_2$Ti$_2$O$_7$. A band of noise, plotted in arbitrary units, is present above the critical frequency, as the monopole gas is too dilute and immobile to completely screen itself.  The monopole hopping time used here is 3 ms and is the only free parameter; the resulting model is qualitatively consistent with all deviations from Arrhenius behavior in our data.}
\end{figure}

Taking the monopole hopping time $\tau_0$ = 3 ms, we plot the full monopole model including the noise and screening terms in Fig.~\ref{monomodel}, in arbitrary units.  We see that the monopole dynamics account not only for the screening at high temperatures and low frequencies, but can also qualitatively account for the non-Arrhenius source of noise as well.  A full modeling of the measured noise spectrum would also require a model for the Arrhenius-like noise behavior.  The defect series that we have measured here suggests that this noise feature is due to defects, most likely stuffed spins.  Previous studies have shown that even nominally stochiometric Ho$_2$Ti$_2$O$_7$ grown by the floating zone method contains approximately 3$\%$ Ho stuffing on the Ti site, or roughly 0.06 stuffed Ho spins per tetrahedron.\cite{Baroudi15}  Given that this exceeds the calculated monopole density at all but the highest temperatures measured (at which the calculated monopole density reaches nearly 0.15 monopoles/tetrahedron), it is unsurprising that defect dynamics would produce magnetic flux noise of a similar magnitude to monopoles.

One possible route towards modeling these dynamics would be to calculate the distribution of activation energies for stuffing defects in a transverse field Ising model.  A holmium atom on a titanium site has 6 nearest neighbor holmium spins that form a closed hexagon in the pyrochlore lattice.  In their normal Ising orientations, these spins each provide an in-plane field for the defect spin.  If the ice state manifold is taken into account in determining the frequency with which different orientations of these nearest neighbor spins will occur, a distribution of activation energies could be calculated.  However, this would not yield the distribution of attempt times which is also necessary for a full accounting of the magnetic dynamics of the defects.  Other defects, such as non-magnetic substitutions on the holmium site or oxygen vacancies,\cite{Sala14} could also give rise to magnetic dynamics that would not be accounted for in this model, and $\tau_0$ could also be temperature dependent for other reasons not considered here, such as non-trivial spin-phonon coupling.\cite{Castelnovo}

\section{\label{sec:HTOConclusion}Conclusion}
We have demonstrated the utility of scanning SQUID magnetic flux spectroscopy by measuring the magnetic flux noise power spectra as a function of temperature in three locations on two samples of the classical spin ice Ho$_2$Ti$_2$O$_7$.  In these measurements, we observe a dominant Arrhenius-like feature that matches the behavior observed in previous bulk ac susceptibility measurements on similar samples.  We identify this feature as the result of the magnetic dynamics of defects in the sample, which we speculate are stuffed spins.

We further identify three qualitative deviations from Arrhenius behavior in all three datasets, namely excess noise below 10 Hz at the lowest temperatures and below 100 Hz at the highest temperatures and screening of the noise at high temperatures and low frequencies.  We find that all three of these behaviors are consistent with the expected dynamics of a dilute, low-mobility gas of magnetic monopoles.  

Our measurements represent a new technique that is complementary to existing magnetic probes used in the study of frustrated magnetic systems. We demonstrate the importance of quantitative modeling for the magnetic dynamics of defects in these systems and the utility of scanning SQUID magnetic flux spectroscopy in disentangling the overlapping magnetic signals of such defects and the essential physics of the system under study, with potential further applications in the study of other, related magnetic systems such as spin liquids.

\begin{acknowledgments}
We thank C. Castelnovo, B. Gaulin, M. Gingras, G. Luke, R. Moessner, H. Noad, J. Rau, K. Ross, and S.L. Sondhi for helpful discussions.  This  work  was primarily supported by the Department of Energy, Office of Science, Basic Energy Sciences, Materials Sciences and Engineering Division, under Contract No. DE-AC02-76SF00515; Ilya Sochnikov was partially supported by the Gordon and Betty  Moore Foundation through grant GBMF3429.  The crystal growth was supported by the US Department of Energy, Division of Basic Energy Sciences, grant DE-SC0019331.
\end{acknowledgments}

\bibliography{mybib} 
\end{document}